\documentstyle[11pt,amssymb]{article}
\def\ba{\begin{eqnarray}}
\def\ea{\end{eqnarray}}

\def\lb{\label}
\def\be{\begin{equation}}
\def\ee{\end{equation}}

\begin{document}

\title{Tri-Sasaki 7-metrics over the QK limit of the Plebanski-Demianski metrics}
\author{O. P. Santillan \thanks{firenzecita@hotmail.com}}
\date {}
\maketitle

\begin{abstract}
We consider a family of conical hyperkahler 8-metrics and we find
the corresponding tri-Sasaki 7-metrics. We find in particular, a
7-dimensional tri-Sasaki fibration over a quaternion Kahler limit of
the Plebanski-Demianski (or AdS-Kerr-Newmann-Taub-Nut) metrics, and
we consider several limits of its parameters. We also find an
squashed version for these metrics, which are of weak $G_2$
holonomy. Construction of supergravity backgrounds is briefly
discussed, in particular examples which do not possess $AdS_4$ near
horizon limit.

\end{abstract}

\section{An sketch of the idea}

   The holographic principle \cite{Klebanov} renewed the interest in constructing 5
and 7-dimensional Einstein manifolds and in particular those
admitting at least one conformal Killing spinor. The number of such
spinors will be related to the number of supersymmetries of the
dual conformal field theory. This leads to consider weak $G_2$ holonomy
spaces, Einstein-Sasaki spaces and tri-Sasaki ones. This spaces
corresponds to $N=1,2,3$ supersymmetries respectively.

   Several of such spaces have been constructed, for instance, in
\cite{Bilal}-\cite{Mateos8}. One of the recent achievements is the
construction of Einstein-Sasaki spaces fibered over orthotoric
Kahler-Einstein spaces \cite{Mateos4}-\cite{orto}. It was indeed
noticed in \cite{orto} that this spaces can be obtained by taking
certain scaling limit of the euclidean Plebanski-Demianski metrics.
In particular, there were found toric Einstein-Sasaki
5-dimensional metrics defined over $S^2\times S^3$.

    The present work deals also with Einstein-Sasaki fibrations over
the euclidean Plebanski-Demianski metric. But we will obtain a
different type of metrics. Our metrics are 7-dimensional instead, and not
only Einstein-Sasaki but tri-Sasaki. The idea behind our
construction is indeed very simple. Our starting point are the Swann
metrics \cite{Swann}, which are \emph{hyperkahler} fibrations over
quaternion Kahler metrics. Their generic form is
$$
g_s=|q|^2g_4+|dq+\omega q|^2,
$$
being $q$ certain quaternion coordinate and $\omega$ an imaginary
quaternion valued 1-form associated to the quaternion Kahler metric $g_4$.
The metric $g_4$ does not depend on $q$, thus under the transformation
$q\to\lambda q$ these metrics are
scaled by a factor $g_s\to\lambda^2 g_s$. It means that they are conical, therefore
they define family tri-Sasaki 7-metrics. We will consider as our base metrics $g_4$ the
euclidean Plebanski-Demianski (or the
AdS-Kerr-Newman-Taub-Nut) metrics which, in some limiting case of the parameters,
are quaternion Kahler \cite{Demianski}. We will extend
them to Swann metrics and finally, we will find
the 7-metric over which $g_8$ is a cone. This will be a tri-Sasaki
metric fibered over the euclidean Plebanski-Demianski 4-metric, which is what
we were looking for.

  We also present an squashed version of these metrics, which are of weak
$G_2$ holonomy. In the last section we construct several 11-dimensional
supergravity backgrounds, some of them with $AdS_4\times X_7$ near horizon limit
and some others without
this property. We achieve this by finding a set of harmonic functions over
the internal Ricci-flat space (the Swann space) which depends not only on
the radius, but also on other angular coordinates of the cone. The key ingredient
in order to construct these functions is to identify the coordinate system
for which the Swann metric takes the generalized Gibbons-Hawking form \cite{Poon}.

\section{Tri-Sasaki fibrations over quaternion Kahler spaces}

\subsection{Quaternion Kahler spaces in brief}

   A quaternion K\"ahler space $M$ is an euclidean $4n$ dimensional
space with holonomy group $\Gamma$ included into the Lie group
$Sp(n)\times Sp(1)\subset SO(4n)$ \cite{Berger}-\cite{Ishihara}.
This affirmation is non trivial if $D>4$, but in $D=4$ there is the
well known isomorphism $Sp(1)\times Sp(1)\simeq SU(2)_L\times
SU(2)_R \simeq SO(4)$ and so to state that $\Gamma\subseteq
Sp(1)\times Sp(1)$ is equivalent to state that $\Gamma\subseteq
SO(4)$. The last condition is trivially satisfied for any oriented
space and gives almost no restrictions, therefore the definition of
quaternion K\"ahler spaces should be modified in $d=4$.

Here we do a brief description of these spaces, more details can be
found in the appendix and in the references therein. For any
quaternion exists three automorphism $J^i$ ($i=1$ ,$2$, $3$) of the
tangent space $TM_x$ at a given point $x$ with multiplication rule
$J^{i} \cdot J^{j} = -\delta_{ij} + \epsilon_{ijk}J^{k}$. The metric
$g_q$ is quaternion hermitian with respect to this automorphism,
that is \be\lb{hermoso} g_q(X,Y)=g(J^i X, J^i Y), \ee being $X$ and
$Y$ arbitrary vector fields. The reduction of the holonomy to
$Sp(n)\times Sp(1)$ implies that the $J^i$ satisfy the fundamental
relation \be\lb{rela2}
\nabla_{X}J^{i}=\epsilon_{ijk}J^{j}\omega_{-}^{k}, \ee being
$\nabla_{X}$ the Levi-Civita connection of $M$ and $\omega_{-}^{i}$
its $Sp(1)$ part. As a consequence of hermiticity of $g$, the tensor
$\overline{J}^{i}_{ab}=(J^{i})_{a}^{c}g_{cb}$ is antisymmetric, and
the associated 2-form
$$
\overline{J}^i=\overline{J}^{i}_{ab} e^a \wedge e^b
$$
satisfies \be\lb{basta}
d\overline{J}^i=\epsilon_{ijk}\overline{J}^{j}\wedge\omega_{-}^{k},
\ee being $d$ the usual exterior derivative.  Corresponding to the
$Sp(1)$ connection we can define the 2-form
$$
F^i=d\omega_{-}^i+\epsilon_{ijk}\omega_{-}^j \wedge \omega_{-}^k.
$$
For any quaternion K\"ahler manifold it follows that \be\lb{lamas}
R^i_{-}=2n\kappa \overline{J}^i, \ee \be\lb{rela} F^i=\kappa
\overline{J}^i, \ee being $\Lambda$ certain constant and $\kappa$
the scalar curvature. The tensor $R^a_{-}$ is the $Sp(1)$ part of
the curvature. The last two conditions implies that $g$ is Einstein
with non zero cosmological constant, i.e, $R_{ij}=3\kappa
(g_{q})_{ij}$ being $R_{ij}$ the Ricci tensor constructed from
$g_q$. The $(0,4)$ and $(2,2)$ tensors
$$
\Theta=\overline{J}^1 \wedge \overline{J}^1 + \overline{J}^2 \wedge
\overline{J}^2 + \overline{J}^3 \wedge \overline{J}^3,
$$
$$
\Xi= J^1 \otimes J^1 + J^2 \otimes J^2 + J^3 \otimes J^3
$$
are globally defined and covariantly constant with respect to the
usual Levi Civita connection for any of these spaces. This implies
in particular that any quaternion K\"ahler space is orientable.

In four dimensions  the K\"ahler triplet $\overline{J}_2$ and the
one forms $\omega^{a}_{-}$ are
$$
\omega^{a}_{-}=\omega^a_{0}- \epsilon_{abc}\omega^b_c,\qquad
\overline{J}_1=e^1\wedge e^2-e^3\wedge e^4,
$$
$$
\overline{J}_2=e^1\wedge e^3-e^4\wedge e^2\qquad
\overline{J}_3=e^1\wedge e^4-e^2\wedge e^3.
$$
In this dimension quaternion K\"ahler spaces are defined by the
conditions (\ref{rela}) and (\ref{lamas}). This definition is
equivalent to state that quaternion K\"ahler spaces are Einstein and
with self-dual Weyl tensor.

In the Ricci-flat limit $\kappa\to 0$ the holonomy of a quaternion
Kahlers space is reduced to a subgroup of $Sp(n)$ and the resulting
spaces are hyperkahler. It follows from (\ref{rela}) and
(\ref{rela2}) that the almost complex structures $J_i$ are
covariantly constant in this case. Also, there exist a frame for
which $\omega_-^i$ goes to zero. In four dimensions this implies
that the spin connection of this frame is self-dual.

\subsection{A general tri-Sasaki family}

   As is well known, any hyperkahler conical metric
define a tri-Sasaki metric by means of $g_8=dr^2+r^2g_7$. A well
known family of conical hyperkahler metrics are the Swann metrics
\cite{Swann}, this are 4n dimensional metrics but we will focus only
in the case $d=8$. The metrics reads \be\lb{Swann2} g_8=|u|^2 g_q +
|du + u \omega_{-}|^2, \ee being $g_q$ any 4-dimensional quaternion
Kahler metric. In the expression for the metric we have defined the
quaternions
$$
u=u_0 + u_1 I + u_2 J + u_3 K ,\;\;\;\;\;\; \overline{u}= u_0 - u_1
I - u_2 J - u_3 K,
$$
and the quaternion one form
$$
\omega_{-}=\omega_{-}^1 I+\omega_{-}^2 J +\omega_{-}^3 K,
$$
constructed with the anti-self-dual spin connection. The
multiplication rule for the quaternions $I$, $J$ and $K$ is deduced
from
$$
I^2=J^2=K^2=-1, \qquad IJ=K, \qquad JI=-K
$$
The metric $g_q$ is assumed to be independent on the coordinates
$u_i$. We easily see that if we scale $u_0$, $u_1$,$u_2$,$u_3$ by
$t>0$ this scales the metric by a homothety $t$, which means that
the metrics (\ref{Swann2}) are conical. Therefore they define a
family of tri-Sasaki metrics, which we will find now. We first
obtain, by defining $\widetilde{u}_i=u_i/u$ that
$$
|du + u \omega_{-}|^2=(du_0-u_i\omega_{-}^i)^2 +(du_i+
u_0\omega_{-}^i + \frac{\epsilon_{ijk}}{2}u_k\omega_{-}^j)^2
$$
$$
=(\widetilde{u}_0 du+ u d\widetilde{u}_0-u
\widetilde{u}_i\omega_{-}^i)^2+(\widetilde{u}_i du+u
d\widetilde{u}_i+u \widetilde{u}_0\omega_{-}^i +
u\frac{\epsilon^{ijk}}{2}\widetilde{u}_j\omega_{-}^k)^2
$$
$$
=du^2+u^2(d\widetilde{u}_0-\widetilde{u}_i\omega_{-}^i)^2+u^2(d\widetilde{u}_i+
\widetilde{u}_0\omega_{-}^i +
\frac{\epsilon_{ijk}}{2}\widetilde{u}_j\omega_{-}^k)^2
$$
$$
+2u u_0 du (d\widetilde{u}_0-\widetilde{u}_i\omega_{-}^i)+2u u_i du
(d\widetilde{u}_i+ \widetilde{u}_0\omega_{-}^i +
\frac{\epsilon^{ijk}}{2}\widetilde{u}_j\omega_{-}^k).
$$
It is not difficult to see that the last two terms are equal to
$$
2u u_0 du (d\widetilde{u}_0-\widetilde{u}_i\omega_{-}^i)+2u u_i du
(d\widetilde{u}_i+ \widetilde{u}_0\omega_{-}^i +
\epsilon_{ijk}\widetilde{u}_i\widetilde{u}_k\omega_{-}^j)=\frac{d(\widetilde{u}_i^2)}{2}
+\frac{\epsilon_{ijk}}{2}\widetilde{u}_i\widetilde{u}_j\omega_{-}^k
$$
But the second term is product of a antisymmetric pseudotensor with
a symmetric expression, thus is zero, and the first term is zero due
to the constraint $\widetilde{u}_i^2=1$. Therefore this calculation
shows that \be\lb{va} |du + u
\omega_{-}|^2=du^2+u^2(d\widetilde{u}_0-\widetilde{u}_i\omega_{-}^i)^2+u^2(d\widetilde{u}_i+
\widetilde{u}_0\omega_{-}^i +
\frac{\epsilon_{ijk}}{2}\widetilde{u}_j\omega_{-}^k)^2. \ee By
introducing (\ref{va}) into (\ref{Swann2}) we find that $g_8$ is a
cone over the following metric \be\lb{trusa}
g_7=g_q+(d\widetilde{u}_0-\widetilde{u}_i\omega_{-}^i)^2+(d\widetilde{u}_i+
\widetilde{u}_0\omega_{-}^i +
\frac{\epsilon_{ijk}}{2}\widetilde{u}_j\omega_{-}^k)^2. \ee This is
the tri-Sasaki metric we were looking for. By expanding the squares
appearing in (\ref{trusa}) we find that
$$
g_7=g_q+(d\widetilde{u}_i)^2+(\omega_{-}^i)^2 +2\omega_{-}^1
(\widetilde{u}_0d\widetilde{u}_1-\widetilde{u}_1d\widetilde{u}_0
+\widetilde{u}_2d\widetilde{u}_3-\widetilde{u}_3d\widetilde{u}_2)
$$
\be\lb{roo} + 2\omega_{-}^2
(\widetilde{u}_0d\widetilde{u}_2-\widetilde{u}_2d\widetilde{u}_0
+\widetilde{u}_2d\widetilde{u}_1-\widetilde{u}_1d\widetilde{u}_3)
+2\omega_{-}^3
(\widetilde{u}_0d\widetilde{u}_3-\widetilde{u}_3d\widetilde{u}_0
+\widetilde{u}_1d\widetilde{u}_2-\widetilde{u}_2d\widetilde{u}_1).
\ee But the expression in parenthesis are a representation of the
$SU(2)$ Maurer-Cartan 1-forms, which are defined by
$$
\sigma_1=-(\widetilde{u}_0d\widetilde{u}_1-\widetilde{u}_1d\widetilde{u}_0
+\widetilde{u}_2d\widetilde{u}_3-\widetilde{u}_3d\widetilde{u}_2)
$$
$$
\sigma_{2}=-(
\widetilde{u}_0d\widetilde{u}_2-\widetilde{u}_2d\widetilde{u}_0
+\widetilde{u}_2d\widetilde{u}_1-\widetilde{u}_1d\widetilde{u}_3)
$$
$$
\sigma_{3}=-
(\widetilde{u}_0d\widetilde{u}_3-\widetilde{u}_3d\widetilde{u}_0
+\widetilde{u}_1d\widetilde{u}_2-\widetilde{u}_2d\widetilde{u}_1).
$$
Therefore the metric (\ref{trusa}) can be reexpressed in simple
fashion as \be\lb{trusa2} g_7=g_q+(\sigma_i-\omega_-^i)^2. \ee This
is one of the expressions that we will use along this work.

    Let us recall that there exist a coordinate system for which the
Maurer-Cartan forms are expressed as \be\lb{mcarta}
\sigma_1=\cos\varphi d\theta+\sin\varphi \sin\theta d\tau,\qquad
\sigma_2=-\sin\varphi d\theta+\cos\varphi \sin\theta d\tau,\qquad
\sigma_3=d\varphi+\cos\theta d\tau. \ee With the help of this
coordinates we will write (\ref{trusa2}) in more customary form for
tri-Sasaki spaces, namely \be\lb{custo} g_7=(d\tau+H)^2+g_6, \ee as
in (\ref{lome}). Here $g_6$ a Kahler-Einstein metric with Kahler
form $\overline{J}$ and $H$ a 1-form such that $dH=2\overline{J}$. A
lengthy algebraic calculation shows that the fiber metric is
$$
(\sigma_i-\omega_-^i)^2=(d\tau+\cos\theta
d\varphi-\sin\theta\sin\varphi\omega_-^1-
\cos\theta\sin\varphi\omega_-^2-\cos\theta\omega_-^3)^2
$$
$$
+(\sin\theta d\varphi-\cos\theta\sin\varphi\omega_-^1-
\cos\theta\cos\varphi\omega_-^2+\sin\theta\omega_-^3)^2
+(d\theta-\sin\varphi\omega_-^2 \cos\varphi\omega_-^1)^2,
$$
from where we read that \be\lb{ache} H=\cos\theta
d\varphi-\sin\theta\sin\varphi\omega_-^1-
\cos\theta\sin\varphi\omega_-^2-\cos\theta\omega_-^3. \ee The
Kahler-Einstein six dimensional metric should be \be\lb{kalonste}
g_6=(\sin\theta d\varphi-\cos\theta\sin\varphi\omega_-^1-
\cos\theta\cos\varphi\omega_-^2+\sin\theta\omega_-^3)^2
+(d\theta-\sin\varphi\omega_-^2 +\cos\varphi\omega_-^1)^2+g_q. \ee
Although we have given the formulas needed for our work, we consider
instructive to give an alternative deduction. This is our next task.
\\

\textit{An alternative deduction of the tri-Sasaki metric
(\ref{trusa2})}
\\

 For any quaternion Kahler space $M$, a linear
combination of the almost complex structures of the form
$J=\widetilde{v}_i J_i$ will be also an almost complex structure on
$M$. Here $\widetilde{v}^i$ denote three "scalar fields"
$\widetilde{v}^i=v^i/v$ being $v=\sqrt{v^iv^i}$. This fields are
supposed to be constant over $M$ and are obviously constrained by
$\widetilde{v}^i \widetilde{v}^i=1$. This means that the bundle of
almost complex structures over $M$ is parameterized by points on the
two sphere $S^2$. This bundle is known as the twistor space $Z$ of
$M$. The space $Z$ is endowed with the metric \be\lb{kahlo}
g_6=\theta_i \theta_i + g_q,\ee where
$\theta_i=d(\widetilde{v}^i)+\epsilon^{ijk}\omega_{-}^j
\widetilde{v}^k$. The constraint $\widetilde{v}^i \widetilde{v}^i=1$
implies that the metric (\ref{kahlo}) is six dimensional. It have
been shown that this metric together with the sympletic two form
\cite{Salomon}, \cite{Lolo} \be\lb{two}
\overline{J}=-\widetilde{u}_i\overline{J}_i
+\frac{\epsilon_{ijk}}{2}\widetilde{v}_i\theta_j\wedge \theta_k, \ee
constitute a \emph{K\"ahler} structure. The calculation of the Ricci
tensor of $g_6$ shows that it is also Einstein, therefore the space
$Z$ is \emph{K\"ahler-Einstein}. The expressions given below are
written for a quaternion Kahler metric normalized such that
$\kappa=1$, for other normalizations certain coefficients must be
included in (\ref{two}). By parameterizing the coordinates
$\widetilde{v}_i$ in the spherical form \be\lb{esferoso}
\widetilde{v}_1=\sin\theta \sin\varphi,\qquad
\widetilde{v}_2=\sin\theta \cos\varphi,\qquad
\widetilde{v}_3=\cos\theta, \ee we find that (\ref{kahlo}) is the
same as (\ref{kalonste}). The isometry group of the Kahler-Einstein
metrics is in general $SO(3)\times G$, being $G$ the isometry group
of the quaternion Kahler basis which also preserve the forms
$\omega_-^i$. The $SO(3)$ part is the one which preserve the
condition $\widetilde{v}_i \widetilde{v}_i=1$. Globally the isometry
group could be larger.

From the definition of Einstein-Sassakian geometry, it follows
directly that the seven dimensional metric \be\lb{owner}
g_7=(d\tau+H)^2+g_6=(d\tau+H)^2+\theta_i \theta_i + g_q, \ee will be
Einstein-Sassaki if $dH=2\overline{J}$, and we need to find an
explicit expression for such $H$. Our aim is to show that this form
is indeed (\ref{ache}). The expression (\ref{two}) needs to be
simplified as follows. We have that
$\theta_i=d(\widetilde{v}^i)+\epsilon^{ijk}\omega_{-}^j
\widetilde{v}^k$. Also by using the condition
$\widetilde{v}_i\widetilde{v}_i=1$ it is found that
$$
\widetilde{v}_i\theta_i=\widetilde{v}_i
d\widetilde{v}_i+\epsilon^{ijk}\widetilde{v}^i\omega_{-}^j
\widetilde{v}^k=\widetilde{v}_i d\widetilde{v}_i=
d(\widetilde{v}_i\widetilde{v}_i)=0.
$$
From the last equality it follows the orthogonality condition
$\widetilde{v}_i\theta_i=0$ which is equivalent to
$$
\theta_3=-\frac{(\widetilde{v}_1\theta_1+\widetilde{v}_2\theta_2)}{\widetilde{v}_3}.
$$
The last relation implies that
$$
\frac{\epsilon_{ijk}}{2}\widetilde{v}_i\theta_j\wedge
\theta_k=\frac{\theta_1\wedge \theta_2}{\widetilde{v}_3}
=\frac{d\widetilde{v}_1\wedge
d\widetilde{v}_2}{\widetilde{v}_3}-d\widetilde{v}_i\wedge
\omega_{-}^i+\frac{\epsilon^{ijk}}{2}\widetilde{v}_i\omega_{-}^j\wedge
\omega_{-}^k.
$$
By another side in a quaternion Kahler manifold with $\kappa=1$ we
always have
$$
\widetilde{J}_i=d\omega_{-}^i+\frac{\epsilon^{ijk}}{2}\omega_{-}^j\wedge
\omega_{-}^k.
$$
Inserting the last two expressions into (\ref{two}) gives a
remarkably simple expression for $\overline{J}$, namely
\be\lb{simple}
\overline{J}=-d(\widetilde{v}_i\omega_{-}^i)+\frac{d\widetilde{v}_1\wedge
d\widetilde{v}_2}{\widetilde{v}_3}. \ee By using (\ref{esferoso}) it
is obtained that
$$
\frac{d\widetilde{v}_1\wedge
d\widetilde{v}_2}{\widetilde{v}_3}=-d\varphi\wedge d\cos\theta.
$$
With the help of the last expression we find that (\ref{simple}) can
be rewritten as
$$
\overline{J}=-d(\widetilde{v}_i\omega_{-}^i)-d\varphi\wedge
d\cos\theta,
$$
from where it is obtained directly that the form $H$ such that
$dH=\overline{J}$ is \cite{yoga} \be\lb{simplon}
H=-\widetilde{v}_i\omega_{-}^i+\cos\theta d\varphi, \ee up to a
total differential term. By introducing (\ref{esferoso}) into
(\ref{simplon}) we find that $H$ is the same than (\ref{ache}), as
we wanted to show.

   It will be of importance for the purposes of the present work to
state these results in a concise proposition.
\\

{\bf Proposition}{ \it Let $g_q$ be a four dimensional Einstein
space with self-dual Weyl tensor, i.e, a quaternion Kahler space. We
assume the normalization $\kappa=1$ for $g_q$. Then the metrics
$$
g_6=(\sin\theta d\varphi-\cos\theta\sin\varphi\omega_-^1-
\cos\theta\cos\varphi\omega_-^2+\sin\theta\omega_-^3)^2
+(d\theta-\sin\varphi\omega_-^2 +\cos\varphi\omega_-^1)^2+g_q,
$$
are Kahler-Einstein whilst
$$
g_7=(\sigma_i-\omega_{-}^i)^2+g_q, \qquad g_8=dr^2+r^2g_7
$$
are tri-Sassakian and hyperkahler respectively. Here $\omega_-^i$ is
the $Sp(1)$ part of the spin connection and $\sigma_i$ are the usual
Maurer-Cartan one forms over $SO(3)$. Moreover the "squashed" family
$$
g_7=(\sigma_i-\omega_{-}^i)^2+5g_q,
$$
is of weak $G_2$ holonomy.}
\\

We will consider the last sentence of this proposition in the next
section. In order to finish this section we would like to describe a
little more the Swann bundles. Under the transformation $u\to G u$
with $G: M \to SU(2)$ the $SU(2)$ instanton $\omega_{-}$ is gauge
transformed as $\omega_{-}\to G\omega_{-}G^{-1}+ GdG^{-1}$.
Therefore the form $du + \omega_{-}u$ is transformed as
$$
du + u \omega_{-}\rightarrow d(Gu) + (G\omega_{-}G^{-1}+ GdG^{-1})
Gu=G du+ (dG+G\omega_{-}-dG)u=G (du + u \omega_{-}),
$$
and it is seen that $du +  \omega_{-}u$ is a well defined
quaternion-valued one form over the chiral bundle. The metric
(\ref{Swann2}) is also well defined over this bundle. Associated to
the metric (\ref{Swann2}) there is a quaternion valued two form
\be\lb{quato} \widetilde{\overline{J}}=u\overline{J}\overline{u}+(du
+ u \omega_{-})\wedge \overline{(du + u \omega_{-})}, \ee and it can
be checked that the metric (\ref{Swann}) is hermitian with respect
to any of the components of (\ref{quato}). Also
$$
d\widetilde{\overline{J}}=du\wedge
(\overline{J}+d\omega_{-}-\omega_{-}\wedge
\omega_{-})\overline{u}+u\wedge
(\overline{J}+d\omega_{-}-\omega_{-}\wedge \omega_{-})d\overline{u}
$$
$$+u(d\overline{J}+\omega_{-}\wedge
d\omega_{-}-d\omega_{-}\wedge \omega_{-})\overline{u}.
$$
The first two terms of the last expression are zero due to
(\ref{rela}). Also by introducing (\ref{rela}) into the relation
(\ref{basta}) it is seen that
$$
d\overline{J}+\omega_{-}\wedge d\omega_{-}-d\omega_{-}\wedge
\omega_{-}=0,
$$
and therefore the third term is also zero. This means that the
metric (\ref{Swann2}) is hyperkahler with respect to the triplet
$\widetilde{\overline{J}}$. The Swann metrics have been considered
in several context in physics, as for instance in
\cite{Suan1}-\cite{Suan4}. It is an important tool also in
differential geometry because the quaternion Kahler quotient
construction correspond to hyperkahler quotients on the Swann
fibrations.

\subsection{The squashed version of the tri-Sasaki family}

   In \cite{Bryant} there were probably constructed
the first examples of $Spin(7)$ holonomy metrics. These examples are
fibered over four dimensional quaternion Kahler metrics defined over
manifold $M$.  This resembles the Swann metrics that we have
presented in (\ref{Swann2}), although the Bryant-Salamon were found
first. The anzatz for the $Spin(7)$ is \be\lb{Swann} g_8=g|u|^2
\overline{g} + f|du + u \omega_{-}|^2, \ee where $f$ and $g$ are two
unknown functions $f(r^2)$ and $g(r^2)$ which will be determined by
the requirement that the holonomy is in $Spin(7)$, i.e, the closure
of the associated 4-form $\Phi_4$. The analogy between the anzatz
(\ref{Swann}) and (\ref{Swann2}) is clear, in fact, if $f=g=1$ the
holonomy will be reduced to $Sp(2)$. A convenient (but not unique)
choice for $\Phi_4$ is the following \be\lb{expresate} \hat{\Phi}= 3
f g [\alpha\wedge \overline{\alpha}\wedge \overline{e}^t \wedge e +
\overline{e}^t \wedge e \wedge \alpha \wedge \overline{\alpha}]+ g^2
\overline{e}^t \wedge e \wedge\overline{e}^t\wedge e + f^2 \alpha
\wedge \overline{\alpha} \wedge \alpha \wedge \overline{\alpha} \ee
where $\alpha = du + u \omega_{-}$. After imposing the condition
$d\Phi_4=0$ to (\ref{Swann}) it is obtained a system of differential
equations for $f$ and $g$ with solution
$$
f=\frac{1}{(2\kappa r^2+c)^{2/5}},
$$
$$
g=(2\kappa r^2+c)^{3/5},
$$
and the corresponding metric \be\lb{sou} g_s=(2\kappa
r^2+c)^{3/5}\overline{g} + \frac{1}{(2\kappa
r^2+c)^{2/5}}|\alpha|^2. \ee Spaces defined by (\ref{sou}) are the
Bryant-Salamon $Spin(7)$ ones. The metrics (\ref{sou}) are non
compact (because $|u|$ is not bounded), and asymptotically conical.
They will be exactly conical only if $c=0$. This is better seen by
introducing spherical coordinates for $u$
$$
u_1=|u| \sin\theta\cos\varphi\cos\tau,
$$
$$
u_2=|u| \sin\theta\cos\varphi\sin\tau,
$$
\be\lb{sos} u_3=|u| \sin\theta\sin\varphi, \ee
$$
u_4=|u| \cos\theta,
$$
and defining the radial variable
$$
r^2=\frac{9}{20}(2\kappa |u|^2 + c)^{3/5}
$$
from which we obtain the spherical form of the metric \be \lb{spo} g
= {dr^2 \over \kappa (1- {c / r^{10/3}})} + {9 \over 100 \, \kappa}
r^2 \left(1 - {c \over r^{10/3}}\right) \left(\sigma^i -
\omega_{-}^i\right)^2 + {9 \over 20} \, r^2 \,\overline{g} \ee being
$\sigma^i$ the left-invariant one-forms on $SU(2)$
$$
\sigma_1=\cos\varphi d\theta + \sin\varphi \sin\theta d\tau
$$
$$
\sigma_2=-\sin\varphi d\theta + \cos\varphi \sin\theta d\tau
$$
$$
\sigma_3=d\varphi + \cos \theta d\tau.
$$
In this case it is clearly seen that (\ref{spo}) are of
cohomogeneity one and thus, by the results presented \cite{Hitchin}-\cite{Bilok},
they define a weak $G_2$ holonomy metric.

Let us fix the normalization $\kappa=1$, as before. Then in the
limit $r>>c$ it is found the behavior \be\lb{cono2} g \approx dr^2 +
r^2\Omega, \ee being $\Omega$ a seven dimensional metric
asymptotically independent of the coordinate $r$, namely
\be\lb{brg2} \Omega= \left(\sigma^i - \omega_{-}^i\right)^2 + 5 g_q
\ee In particular the subfamilies of (\ref{spo}) with $c=0$ are
exactly conical and their angular part is (\ref{brg2}). This seven
dimensional metric is of weak $G_2$ holonomy and possesses an
$SO(3)$ isometry action associated with the $\sigma^i$. If also the
four dimensional quaternion Kahler metric has an isometry group $G$
that preserve the $\omega_{-}^i$, then the group is enlarged to
$SO(3)\times G$.

This metric can also be obtained by introducing four new coordinates
$(r, \theta, \phi, \varphi)$ and using an anzatz of the form
\be\lb{uns} g= \alpha^2 dr^2 +\beta^2
(\sigma_i-A_i)^2+\gamma^2\overline{g}, \ee being $\sigma_i$ the
$SU(2)$ left-invariant one forms. The unknown functions $\alpha$,
$\beta$ and $\gamma$ are supposed only functions of $r$. By
requiring $d\hat{\Phi}=0$ for the four-form constructed with
\be\lb{vacia} e´^0=\alpha dr,\qquad e´^i=\beta(\sigma_i-A_i),\qquad
e´^a=\gamma e^a, \ee the metrics (\ref{spo}) will be obtained again.

\subsection{A test of the formulas}

    It is extremely important to compare the weak $G_2$ holonomy metrics
(\ref{brg2}) and the tri-Sasaki metrics (\ref{trusa2}). The only
difference between the two metrics is a factor 5 in front of $g_q$
in (\ref{brg2}). Both metrics possess the same isometry group. At
first sight it sounds possible to absorb this factor 5 by a simple
rescale of the coordinates and therefore to conclude that both
metrics are the same. \emph{But this is not true}. We are fixing the
normalization $\kappa=1$ in both cases, thus this factor should be
absorbed only by an rescaling on the coordinates of the fiber. There
is no such rescaling. Therefore, due to this factor 5, both metrics
are different. This is what one expected, since they are metrics of
different type.

    We can give an instructive example to understand why this is so.
With this purpose in mind let us consider the Fubini-Study metric on
CP(2). There exists a coordinate system for which the metric take
the form \be\lb{fubi}
g_f=2d\mu^2+\frac{1}{2}\sin^2\mu\widetilde{\sigma}_3^2
+\frac{1}{2}\sin^2\mu\cos^2\mu(\widetilde{\sigma}_1^2+\widetilde{\sigma}_2^2).
\ee We have denoted the Maurer-Cartan one-forms of this expression
as $\widetilde{\sigma}_i$ in order to not confuse them with the
$\sigma_i$ appearing in (\ref{brg2}) and (\ref{trusa2}). The
anti-self-dual part of the spin connection is \be\lb{rolon}
\omega_-^1=-\cos\mu\widetilde{\sigma}_1, \qquad
\omega_-^2=\cos\mu\widetilde{\sigma}_2, \qquad
\omega_-^3=-\frac{1}{2}(1+\cos\mu)\widetilde{\sigma}_3. \ee The two
metrics that we obtain by use of (\ref{brg2}) and (\ref{trusa2}) are
\be\lb{tro} g_7=
2b\;d\mu^2+\frac{1}{2}\sin^2\mu\widetilde{\sigma}_3^2
+b\;\frac{1}{2}\sin^2\mu\cos^2\mu(\widetilde{\sigma}_1^2+b\;\widetilde{\sigma}_2^2)\ee
$$
+(\sigma_1+\cos\mu\widetilde{\sigma}_1)^2+(\sigma_2-\cos\mu\widetilde{\sigma}_2)^2+
(\sigma_3+\frac{1}{2}(1+\cos\mu)\widetilde{\sigma}_3)^2.
$$
If (\ref{brg2}) and (\ref{trusa2}) are correct, then $b=1$
corresponds to a tri-Sasaki metric and $b=5$ to a weak $G_2$
holonomy one. This is true. Locally this metrics are the same that
$N(1,1)_I$ and $N(1,1)_{II}$ given in \cite{Donpa}, which are known
to be tri-Sasaki and weak $G_2$. We see therefore that this number
five in front of the quaternion Kahler metric is relevant and change
topological properties of the metric (such as the number of
conformal Killing spinors). In other words, we will construct in
this work an infinite doublet of 7-dimensional metrics, one with one
conformal Killing spinor and other with three.

\section{Explicit tri-Sasaki metrics over manifolds and orbifolds}

\subsection{Quaternion Kahler limit of AdS-Kerr-Newman-Taub-Nut}

In this subsection we will describe certain toric quaternion Kahler
orbifolds which are obtained by a Wick rotation of the Plebanski and
Demianski solution \cite{Demianski}. In fact, these spaces has been
discussed in detail in \cite{Demianski2}-\cite{Demianski6}. The
distance element is \be\lb{plebanski} g_q=\frac{x^2-y^2}{P}dx^2+
\frac{x^2-y^2}{Q}dy^2+\frac{P}{x^2-y^2}(d\alpha+y^2d\beta)^2+
\frac{Q}{x^2-y^2}(d\alpha+x^2d\beta)^2 \ee being $P(x)$ and $Q(y)$
polynomials of the form
$$
P(x)=q-2 s x-t x^2- \kappa x^4,\qquad Q(y)=-P(y),
$$
being $(q,s,t,\kappa)$ four parameters. These expressions can be
rewriten as
$$
P(x)=-\kappa(x-r_1)(x-r_2)(x-r_3)(x-r_4),\qquad Q(y)=-P(y),
$$
$$
r_1+r_2+r_3+r_4=0,
$$
the last condition comes from the fact that $P(x)$ contains no cubic
powers of $x$. The two commuting Killing vectors are
$\partial_{\alpha}$ and $\partial_{\beta}$.

    The metric (\ref{plebanski})
is invariant under the transformation $x\leftrightarrow y$. The
transformations $x\rightarrow -x$, $y\rightarrow -y$,
$r_i\rightarrow -r_i$ are also a symmetry of the metric. In addition
the symmetry $(x, y, \alpha, \beta)\rightarrow (\lambda x, \lambda
y, \frac{\alpha}{\lambda}, \frac{\beta}{\lambda^3})$,
$r_i\rightarrow \lambda r_i$ can be used in order to put one
parameter equal to one, so there are only three effective parameters
here. The domains of definition are determined by
$$
(x^2-y^2)P(x)\geq 0, \qquad (x^2-y^2)Q(y)\geq 0.
$$
The anti-self-dual part of the spin connection is
$$
\omega_-^1=\frac{\sqrt{PQ}}{y-x}d\beta, \qquad
\omega_-^3=\frac{1}{x-y}\bigg(\sqrt{\frac{Q}{P}}dx+\sqrt{\frac{P}{Q}}dy\bigg),
$$
\be\lb{party}
\omega_-^2=-\kappa(x-y)d\alpha+\frac{1}{x-y}\bigg(q-s(x+y)-t xy-
\kappa x^2 y^2\bigg)d\beta, \ee  (see for instance
\cite{Mahapatrah}). We will need (\ref{party}) in the following.
These solutions are the self-dual limit of the
AdS-Kerr-Newmann-Taub-Nut solutions, the last one corresponds to the
polynomials
$$
P(x)=q-2 s x-t x^2-\kappa x^4,\qquad Q(y)=-q+2 s' x+t x^2+\kappa
x^4,
$$
and are always Einstein, but self-dual if and only if $s'=s$. We
will concerned with this limit in the following, because is the one
which is quaternion Kahler. If we define the new coordinates
$$
y=\widetilde{r},\qquad x=a \cos\widetilde{\theta}+N,
$$
$$
\alpha=t+(\frac{N^2}{a}+a)\frac{\widetilde{\phi}}{\Xi},\qquad
\beta=-\frac{\widetilde{\phi}}{a\Xi},
$$
where we have introduced the parameters
$$
\Xi=1-\kappa a^2,\qquad q=-a^2+N^2(1-3\kappa a^2+3\kappa N^2),
$$
$$
s=N(1-\kappa a^2+4N^2), \qquad t=-1-\kappa a^2-6 \kappa N^2,
$$
then the functions $P$ and $Q$ are expressed as
$$
P=-a^2\sin^2\widetilde{\theta}[1-\kappa(4aN
\cos\widetilde{\theta}+a^2\cos^2\widetilde{\theta})],
$$
$$
Q=-\widetilde{r}^2-N^2+2s'\widetilde{r}+a^2+\kappa(\widetilde{r}^4-a^2\widetilde{r}^2-6\widetilde{r}^2N^2+3a^2N^2-3N^4),
$$
and the metric take the AdS-Kerr-Newman-Taub-Nut form
$$
g_q=\frac{R^2}{1-\kappa(a^2\cos^2\widetilde{\theta}+4aN
\cos\widetilde{\theta})}d\widetilde{\theta}^2+\frac{R^2}{\lambda^2}dr^2
+\frac{\lambda^2}{R^2}[d\widetilde{t}+(\frac{a\sin^2\widetilde{\theta}}{\Xi}
-\frac{2N\cos\widetilde{\theta}}{\Xi})d\widetilde{\phi}]^2
$$
\be\lb{demi}
+\frac{\sin^2\widetilde{\theta}[1-\kappa(a^2\cos^2\widetilde{\theta}+4aN
\cos\widetilde{\theta})]}{R^2}[a
d\widetilde{t}-\frac{r^2-a^2-N^2}{\Xi}d\widetilde{\phi}]^2, \ee
being $R$ and $\lambda$ defined by
$$
R=\widetilde{r}^2-(a\cos\widetilde{\theta}+N)^2,
$$
$$
\lambda=\widetilde{r}^2+N^2-2s'\widetilde{r}-a^2-\kappa(\widetilde{r}^4-a^2\widetilde{r}^2-6\widetilde{r}^2N^2+3a^2N^2-3N^4).
$$
Notice that the self-dual limit corresponds to the choice
$s'=N(1-\kappa a^2+4aN^2)$ in all the expressions. The parameter
$\kappa$ is the scalar curvature of the metric and we fix
$\kappa=1$, as we did previously.

   These metrics have interesting limits. For $a=0$ and $N$
different from zero becomes the AdS Taub-Nut solution with local
metric \be\lb{ADSTN} g_q=V(\widetilde{r})(d\widetilde{t}-2N
\cos\widetilde{\theta}
d\widetilde{\phi})^2+\frac{d\widetilde{r}^2}{V(\widetilde{r})}+(\widetilde{r}^2-N^2)(d\widetilde{\theta}^2+\sin^2\widetilde{\theta}
d\widetilde{\phi}^2), \ee being $V(\widetilde{r})$ given by
$$
V(\widetilde{r})=\frac{\lambda}{R^2}=\frac{1}{\widetilde{r}^2-N^2}\bigg(\widetilde{r}^2+N^2
-(\widetilde{r}^4-6N^2\widetilde{r}^2-3N^4)-2s'\widetilde{r}\bigg).
$$
This metric has been considered in different context
\cite{Demianski2}-\cite{Demianski6}. The parameter $s'$ is a mass
parameter and $N$ is a nut charge. Both parameters are not
independent in the quaternion Kahler limit, in fact the self-duality
condition $s'=s$ relates them as $s'=N(1+4N^2)$. If the mass were
arbitrary then the metric will possess a "bolt", but in this case
the metric will possess a "nut", that is, a zero dimensional regular
fixed point set. The isometry group of this metric is enhanced from
$U(1)\times U(1)$ to $SU(2)\times U(1)$ in this limit. The anti
self-dual part of the spin connection reads
$$
\omega_-^1=-\sqrt{(\widetilde{r}+N)V(\widetilde{r})}\sin
\widetilde{\theta}d\widetilde{\phi}, \qquad
\omega_-^3=\sqrt{(\widetilde{r}+N)V(\widetilde{r})}d\widetilde{\theta},
$$
\be\lb{rn}
\omega_-^2=(\widetilde{r}-N)d\widetilde{t}+g(\widetilde{r})\cos\widetilde{\theta}d\widetilde{\phi},
\ee being $g(\widetilde{r})$ defined by
$$
g(\widetilde{r})=\bigg(\frac{N^2(\widetilde{r}-N)+N(1+4N^2)+(1+6N^2)
\widetilde{r}- 2N \widetilde{r}^2}{\widetilde{r}-N}\bigg).
$$
By taking the further limit $N=0$, that is, but switching off the
mass and the charge, we obtain after introducing the new radius
$\widetilde{r}=\sin\widetilde{\rho}$ the following distance element
$$
g_q=\cos^2\widetilde{\rho}d\widetilde{t}^2+d\widetilde{\rho}^2+\sin^2\widetilde{\rho}(d\widetilde{\theta}^2+\sin^2\widetilde{\theta}
d\widetilde{\phi}^2).
$$
The anti-self-dual spin connection takes the simple form
$$
\omega_-^1=\cos\widetilde{\rho}\sin
\widetilde{\theta}d\widetilde{\phi}, \qquad
\omega_-^2=\sin\widetilde{\rho}d\widetilde{t}+\cos\widetilde{\theta}d\widetilde{\phi},\qquad
\omega_-^3=\cos\widetilde{\rho}d\widetilde{\theta},
$$
and it follows that we have obtained the metric of the sphere
$S^4=SO(5)/SO(4)$. If we would choose negative scalar curvature
instead, this limit would correspond to the non compact space
$SO(4,1)/SO(4)$. Both cases are maximally symmetric and for this
reason this is called the $AdS_4$ limit of the AdS-Taub-Nut
solution. The only known quaternion kahler manifolds in 4-dimensions
are the $S^4$ and CP(2). The CP(2) manifold limit is obtained by
defining the new coordinates $\widehat{r}=N(\widetilde{r}-N)$ and
$\widetilde{t}=2N\xi$ and taking the limit $N\to \infty$. The
result, after defining
$\widetilde{\rho}=\widehat{r}^2/4(1+\widehat{r}^2)$, is the metric
\be\lb{sepedos}
g_q=\frac{\widetilde{\rho}^2}{2(1+\widetilde{\rho}^2)^2}(d\xi-
\cos\widetilde{\theta}
d\widetilde{\phi})^2+\frac{2d\widetilde{\rho}^2}{(1+\widetilde{\rho}^2)^2}
+\frac{\widetilde{\rho}^2}{2(1+\widetilde{\rho}^2)^2}(d\widetilde{\theta}^2+\sin^2\widetilde{\theta}
d\widetilde{\phi}^2). \ee By noticing that $\sigma_3=d\xi-
\cos\widetilde{\theta} d\widetilde{\phi}$ and that
$\sigma_1^2+\sigma_2^2=d\widetilde{\theta}^2+\sin^2\widetilde{\theta}
d\widetilde{\phi}^2$ we recognize from (\ref{sepedos}) the Bianchi
IX form for the Fubbini-Study metric on CP(2)=$SU(3)/SU(2)$.
Instead, if we put $N=0$ in (\ref{demi}), the result will be the
AdS-Kerr euclidean solution in Boyer-Linquidst coordinates, namely
\be\lb{demi}
g_q=\frac{\widetilde{r}^2-a^2\cos\widetilde{\theta}^2}{1-a^2\cos^2\widetilde{\theta}}
d\widetilde{\theta}^2+\frac{\widetilde{r}^2-a^2\cos\widetilde{\theta}^2}{(\widetilde{r}^2-a^2)(1-\widetilde{r}^2)}d\widetilde{r}^2
+\frac{(\widetilde{r}^2-a^2)(1-\widetilde{r}^2)}{\widetilde{r}^2-a^2\cos\widetilde{\theta}^2}(\;d\widetilde{t}+\frac{a\sin^2\widetilde{\theta}}{\Xi}
d\widetilde{\phi}\;)^2\ee
$$
+\frac{\sin^2\widetilde{\theta}(1-a^2\cos^2\widetilde{\theta})}{\widetilde{r}^2-a^2\cos^2\widetilde{\theta}}
(\;a
d\widetilde{t}-\frac{\widetilde{r}^2-a^2}{\Xi}d\widetilde{\phi}\;)^2.$$
The anti-self-dual connection $\omega_-^i$ is in this case $$
\omega_-^1=-\frac{1}{\widetilde{r}-a\cos\widetilde{\theta}}\sqrt{(1-a^2\cos\widetilde{\theta}^2)
(\widetilde{r}^2-a^2)(1-\widetilde{r}^2)}
\frac{\sin\widetilde{\theta}}{\Xi}d\widetilde{\phi},
$$
$$
\omega_-^2=(\widetilde{r}-a\cos\widetilde{\theta})d\widetilde{t}+\frac{1}{\widetilde{r}-a\cos\widetilde{\theta}}
\frac{W(\widetilde{r}, \widetilde{\theta})}{\Xi}d\widetilde{\phi},
$$
\be\lb{rn2}
\omega_-^3=\frac{1}{\widetilde{r}-a\cos\widetilde{\theta}}\bigg(
\sqrt{\frac{(\widetilde{r}^2-a^)(1-\widetilde{r}^2)}{1-a^2\cos^2\widetilde{\theta}}}d\widetilde{\theta}
-\sqrt{\frac{1-a^2\cos^2\widetilde{\theta}}{(\widetilde{r}^2-a^2)(1-\widetilde{r}^2)}}a\sin{\widetilde{\theta}}d\widetilde{r}\bigg),
\ee where we have defined the function \be\lb{sh}
W(\widetilde{r},\widetilde{\theta})=[(\widetilde{r}-a\cos\widetilde{\theta})^2
-a+(1+a^2)\widetilde{r} \cos\widetilde{\theta}-a
\widetilde{r}^2\cos^2\widetilde{\theta}]. \ee The parameter $a$ is
usually called rotational parameter, although we have no the notion
of rotational black hole in euclidean signature. The mass parameter
$s$ and the nut charge are zero in this case.

\subsection{Tri-Sassaki and weak $G_2$ over AdS-Kerr and AdS-Taub-Nut}

  We are now in position to find new compact
tri-Sasaki and weak $G_2$ holonomy metrics. The main ingredient in
this construction is the proposition 1, applied to limiting cases of
the euclidean Plebanski-Demianski solution (\ref{demi}). Let us turn our attention to
this task.
\\

\textit{The AdS-Taub-Nut case}
\\

It is direct, by using proposition 1 and the lifting formula
(\ref{brg2}), to work out tri-Sassaki and weak $G_2$ holonomy
metrics fibered over the AdS-Taub-Nut metrics (\ref{ADSTN}), the
result is
$$
g_7=(\sqrt{(\widetilde{r}+N)V(\widetilde{r})}\sin
\widetilde{\theta}d\widetilde{\phi}+\sigma_1)^2+
\bigg((\widetilde{r}-N)d\widetilde{t}+g(\widetilde{r})\cos\widetilde{\theta}d\widetilde{\phi}-\sigma_2\bigg)^2
$$
$$
+(\sqrt{(\widetilde{r}+N)V(\widetilde{r})}d\widetilde{\theta}-\sigma_3)^2+b\;\bigg(\;V(\widetilde{r})(d\widetilde{t}-2N
\cos\widetilde{\theta}
d\widetilde{\phi})^2+\frac{d\widetilde{r}^2}{V(\widetilde{r})}+(\widetilde{r}^2-N^2)(d\widetilde{\theta}^2+\sin^2\widetilde{\theta}
d\widetilde{\phi}^2)\bigg).
$$
Although the base quaternion Kahler space possess $SU(2)\times U(1)$
isometry, this group does not preserve the fibers, so the isometry
group is $SU(2)\times U(1)^2$, being the $SU(2)$ group related to
the Maurer-Cartan forms of the fiber metric and $U(1)^2$ generated
by $\partial_{\widetilde{t}}$ and $\partial_{\widetilde{\phi}}$. Let
us notice that we have a third commuting Killing vector, which is
the Reeb vector $\partial_{\tau}$, which is present in the
expression for the Maurer-Cartan forms $\sigma_i$. Therefore we have
a $T^3$ subgroup of isometries. By taking into account the explicit
form of the $\sigma_i$´s given in (\ref{mcarta}) we obtain the
following metric components
$$
g_{\widetilde{t}\widetilde{t}}=(\widetilde{r}-N)^2+ b
V(\widetilde{r}), \qquad g_{\widetilde{\phi}\widetilde{\phi}}=4b N^2
V(\widetilde{r})\cos^2\widetilde{\theta}
+(\widetilde{r}^2-N^2)\sin^2\widetilde{\theta}
$$
$$
g_{\widetilde{\theta}\widetilde{\theta}}=b(\widetilde{r}^2-N^2)+(\widetilde{r}+N)V(\widetilde{r}),
\qquad g_{\widetilde{r}\widetilde{r}}=\frac{b}{V(\widetilde{r})},
\qquad g_{\tau\tau}= g_{\varphi\varphi}=g_{\theta\theta}=1
$$
$$
g_{\widetilde{t}\widetilde{\phi}}=- 2N b
V(\widetilde{r})\cos\widetilde{\theta}+(\widetilde{r}-N)g(\widetilde{r})\cos\widetilde{\theta},
\qquad g_{\widetilde{t}\tau}=-(\widetilde{r}-N)\cos\varphi
\sin\theta
$$
\be\lb{azere}
g_{\widetilde{\phi}\tau}=\sqrt{(\widetilde{r}+N)V(\widetilde{r})}\sin\widetilde{\theta}
\sin\theta \sin\varphi+g(\widetilde{r})\cos\widetilde{\theta}
\sin\theta \cos\varphi \ee
$$
g_{\widetilde{\phi}\theta}=\sqrt{(\widetilde{r}+N)V(\widetilde{r})}\sin\widetilde{\theta}
\cos\varphi+g(\widetilde{r})\cos\widetilde{\theta} \sin\varphi
$$
$$
g_{\widetilde{\theta}\tau}=-\sqrt{(\widetilde{r}+N)V(\widetilde{r})}\cos\theta,
\qquad
g_{\widetilde{\theta}\varphi}=-\sqrt{(\widetilde{r}+N)V(\widetilde{r})}
$$
$$
g_{\widetilde{t}\theta}=-(\widetilde{r}-N)\sin\varphi
\sin\theta,\qquad g_{\tau\varphi}=\cos\theta,
$$
the remaining components are all zero. The parameter $b$ take the
values 1 or 5, $b=1$ corresponds to an Einstein-Sassaki metric,
while $b=5$ corresponds to a weak $G_2$ holonomy metric.
\\

\textit{The AdS-Kerr-Newmann case}
\\

For the rotating case, that is, for the AdS-Kerr-Newman metrics
(\ref{demi}) we obtain the metrics
\be\lb{demil}
g_q=\bigg(\frac{\sqrt{f(\widetilde{\theta})
c(\widetilde{r})d(\widetilde{r})}}{e(\widetilde{r},
\widetilde{\theta})}
\frac{\sin\widetilde{\theta}}{\Xi}d\widetilde{\phi}-\sigma_1\bigg)^2+\bigg(e(\widetilde{r},
\widetilde{\theta})d\widetilde{t}+ \frac{W(\widetilde{r},
\widetilde{\theta})}{\Xi e(\widetilde{r},
\widetilde{\theta})}d\widetilde{\phi} -\sigma_2\bigg)^2 \ee
$$
+\bigg(
\sqrt{\frac{c(\widetilde{r})d(\widetilde{r})}{f(\widetilde{\theta})}}\frac{d\widetilde{\theta}}{e(\widetilde{r},
\widetilde{\theta})}
-\sqrt{\frac{f(\widetilde{\theta})}{c(\widetilde{r})d(\widetilde{r})}}\frac{a\sin{\widetilde{\theta}}}{e(\widetilde{r},
\widetilde{\theta})}d\widetilde{r} -\sigma_3\bigg)^2+\frac{b
\;f(\widetilde{\theta}) \sin^2\widetilde{\theta}
}{\widetilde{r}^2-a^2\cos^2\widetilde{\theta}} (\;a
d\widetilde{t}-\frac{c(\widetilde{r})}{\Xi}d\widetilde{\phi}\;)^2
$$
$$
+ \frac{b\;
c(\widetilde{r})d(\widetilde{r})}{\widetilde{r}^2-a^2\cos\widetilde{\theta}^2}(\;d\widetilde{t}
+\frac{a\sin^2\widetilde{\theta}}{\Xi}
d\widetilde{\phi}\;)^2+\frac{\widetilde{r}^2-a^2\cos\widetilde{\theta}^2}{f(\widetilde{\theta})}
b\;d\widetilde{\theta}^2+\frac{\widetilde{r}^2-a^2\cos\widetilde{\theta}^2}{c(\widetilde{r})d(\widetilde{r})}b
\;d\widetilde{r}^2 ,
$$
where we have introduced the functions
$$
f(\widetilde{\theta})=1-a^2\cos^2\widetilde{\theta},\qquad
c(\widetilde{r})=\widetilde{r}^2-a^2, \qquad
d(\widetilde{r})=1-\widetilde{r}^2\qquad e(\widetilde{r},
\widetilde{\theta})=\widetilde{r}-a\cos\widetilde{\theta}.
$$
The local isometry is $SU(2)\times U(1)^2$ and as before, the
vectors $\partial_{\widetilde{t}}$,  $\partial_{\widetilde{\phi}}$.
and $\partial_{\tau}$ generate a $T^3$ isometry subgroup. From
expression (\ref{demil}) we read the following components
$$
g_{\widetilde{t}\widetilde{t}}=\frac{b\;
c(\widetilde{r})d(\widetilde{r})}{\widetilde{r}^2-a^2\cos\widetilde{\theta}^2}
+a^2\frac{b \;f(\widetilde{\theta}) \sin^2\widetilde{\theta}
}{\widetilde{r}^2-a^2\cos^2\widetilde{\theta}} +e^2(\widetilde{r},
\widetilde{\theta})
$$
$$
g_{\widetilde{\phi}\widetilde{\phi}}=\frac{b\;
c(\widetilde{r})d(\widetilde{r})}{\widetilde{r}^2-a^2\cos\widetilde{\theta}^2}\frac{a^2\sin^4\widetilde{\theta}}{\Xi^2}
+\frac{b \;f(\widetilde{\theta}) \sin^2\widetilde{\theta}
}{\widetilde{r}^2-a^2\cos^2\widetilde{\theta}}\frac{c^2(\widetilde{r})}{\Xi^2}
+\frac{W^2(\widetilde{r}, \widetilde{\theta})}{\Xi^2
e^2(\widetilde{r}, \widetilde{\theta})} +\frac{f(\widetilde{\theta})
c(\widetilde{r})d(\widetilde{r})}{e^2(\widetilde{r},
\widetilde{\theta})} \frac{\sin^2\widetilde{\theta}}{\Xi^2}
$$
$$
g_{\widetilde{\theta}\widetilde{\theta}}=\frac{\widetilde{r}^2-a^2\cos\widetilde{\theta}^2}{f(\widetilde{\theta})}
b+\frac{1}{e^2(\widetilde{r},
\widetilde{\theta})}\frac{c(\widetilde{r})d(\widetilde{r})}{f(\widetilde{\theta})},
\qquad
g_{\widetilde{r}\widetilde{r}}=\frac{\widetilde{r}^2-a^2\cos\widetilde{\theta}^2}{c(\widetilde{r})d(\widetilde{r})}b
+\frac{f(\widetilde{\theta})}{c(\widetilde{r})d(\widetilde{r})}\frac{a^2\sin^2{\widetilde{\theta}}}{e^2(\widetilde{r},
\widetilde{\theta})},
$$
$$
g_{\widetilde{r}\widetilde{\theta}}=-\frac{a\sin{\widetilde{\theta}}}{e^2(\widetilde{r},
\widetilde{\theta})}, \qquad
g_{\widetilde{r}\varphi}=\sqrt{\frac{f(\widetilde{\theta})}{c(\widetilde{r})d(\widetilde{r})}}
\frac{a\sin{\widetilde{\theta}}}{e(\widetilde{r},
\widetilde{\theta})} \qquad
g_{\widetilde{r}\tau}=\sqrt{\frac{f(\widetilde{\theta})\cos\theta}{c(\widetilde{r})d(\widetilde{r})}}\frac{a\sin{\widetilde{\theta}}\cos\theta}{e(\widetilde{r},
\widetilde{\theta})}
$$
$$
g_{\widetilde{t}\widetilde{\phi}}= \frac{b\;
c(\widetilde{r})d(\widetilde{r})}{\widetilde{r}^2-a^2\cos\widetilde{\theta}^2}\frac{a\sin^2\widetilde{\theta}}{\Xi}
+a\frac{b \;f(\widetilde{\theta}) \sin^2\widetilde{\theta}
}{\widetilde{r}^2-a^2\cos^2\widetilde{\theta}}
\frac{c(\widetilde{r})}{\Xi} +\frac{W(\widetilde{r},
\widetilde{\theta})}{\Xi},
$$
$$
g_{\tau\tau}= g_{\varphi\varphi}=g_{\theta\theta}=1, \qquad
g_{\widetilde{t}\tau}=-e(\widetilde{r},\widetilde{\theta})\cos\varphi
\sin\theta
$$
\be\lb{azere2}
g_{\widetilde{\phi}\tau}=-\frac{\sqrt{f(\widetilde{\theta})
c(\widetilde{r})d(\widetilde{r})}}{e(\widetilde{r},
\widetilde{\theta})} \frac{\sin\widetilde{\theta}}{\Xi} \sin\theta
\sin\varphi-\frac{W(\widetilde{r}, \widetilde{\theta})}{\Xi
e(\widetilde{r}, \widetilde{\theta})} \sin\theta \cos\varphi \ee
$$
g_{\widetilde{\phi}\theta}=\frac{W(\widetilde{r},
\widetilde{\theta})}{\Xi e(\widetilde{r}, \widetilde{\theta})}
\sin\varphi+\frac{\sqrt{f(\widetilde{\theta})
c(\widetilde{r})d(\widetilde{r})}}{e(\widetilde{r},
\widetilde{\theta})} \frac{\sin\widetilde{\theta}}{\Xi}\cos\varphi
$$
$$
g_{\widetilde{\theta}\tau}=
-\sqrt{\frac{c(\widetilde{r})d(\widetilde{r})}{f(\widetilde{\theta})}}\frac{\cos\theta}{e(\widetilde{r},
\widetilde{\theta})}, \qquad
g_{\widetilde{\theta}\varphi}=-\sqrt{\frac{c(\widetilde{r})d(\widetilde{r})}{f(\widetilde{\theta})}}\frac{1}{e(\widetilde{r},
\widetilde{\theta})}
$$
$$
g_{\widetilde{t}\theta}=-e(\widetilde{r},
\widetilde{\theta})\sin\varphi,\qquad g_{\tau\varphi}=\cos\theta
$$
and the other components are zero.
\\

\textit{The manifold limit}
\\

We saw that the $S^4$ and CP(2) metrics are limits of (\ref{ADSTN}).
By taking these limits, a doublet of tri-Sasaki and weak $G_2$
metric fibered over these manifolds arise from (\ref{azere}). For
$S^4$ the result is the following metric
$$
g_7=(\sin\widetilde{\rho}d\widetilde{t}+\cos\widetilde{\theta}d\widetilde{\phi}+\sin\varphi
d\theta -\cos\varphi \sin\theta d\tau)^2 +
(\cos\widetilde{\rho}d\widetilde{\theta}-d\varphi - \cos \theta
d\tau)^2$$ \be\lb{chacha} +(\cos\widetilde{\rho}\sin
\widetilde{\theta}d\widetilde{\phi}-\cos\varphi d\theta -\sin\varphi
\sin\theta
d\tau)^2+b\cos^2\widetilde{\rho}d\widetilde{t}^2+bd\widetilde{\rho}^2+b\sin^2\widetilde{\rho}(d\widetilde{\theta}^2+\sin^2\widetilde{\theta}
d\widetilde{\phi}^2) . \ee The expression for the tri-Sasaki 7-metrics fibered
over CP(2) was already given in (\ref{tro}), it is well known and
have been considered already, so we will not discuss it again.

\subsection{AdS and non AdS backgrounds from harmonic functions}

In the context of the $AdS/CFT$  correspondence \cite{Juanma}-\cite{Klebanov} there
is of
interest to construct eleven dimensional supergravity backgrounds of
the form
\be\lb{genero2}
g_{11}=H^{-2/3}(-dt^2+dx^2+dy^2)+
H^{1/3}(dr^2+r^2 g_{7}), \ee
$$
F=\pm dx\wedge dy\wedge dt\wedge dH^{-1}
$$
where the conical metric $g_8=dr^2+r^2 g_{7}$ is Ricci flat and $H$
is an harmonic function over the space $M_8$ where $g_8$ is defined.
Usually one consider radial harmonic functions given by
$$
H(r)=1+\frac{2^5\pi^2N l_p^6}{r^6}.
$$
This solution describe $N$ M2 branes. The near horizon limit of this
geometry is obtained taking the 11 dimensional Planck length $l_p\to
0$ and keeping fixed $U=r^2/l_p^3$. The resulting background is
$AdS_4\times X_7$, being $X_7$ is an Einstein manifold with
cosmological constant $\Lambda=5$, and the radius of $AdS_4$ is
$2R_{AdS}=l_p(2^5\pi^2 N)^{1/6}$. Such solutions have the generic
form \be\lb{solu} g_{11}=g_{AdS}+g_7, \qquad F_4\sim \omega_{AdS},
\ee being $\omega_{AdS}$ the volume form of $AdS_4$.  This
backgrounds are in general associated to three dimensional conformal
field theories arising as the infrared limit of the world volume
theory of N coincident M2 branes located a the singularity of
$M_3\times X_8$. Also in this case, the number of supersymmetries of
the field theory is determined by the holonomy of $X_8$. In the case
of $Spin(7)$, $SU(4)$ or $Sp(2)$ holonomies we have $N=1,2,3$
supersymmetries, respectively. This implies that the 7-dimensional
cone will be of weak $G_2$ holonomy (if the eight dimensional metric
is of cohomogenity one, see below), tri-Sassaki or a
Sassaki-Einstein, respectively. If $g_8$ is flat, then we have the
maximal number of supersymmetries, namely eight.

   Non $AdS_4$ backgrounds are also of interest because they give
rise to non conformal field theory duals. Therefore it is of
interest to find harmonic functions which are functions not only of
the radius $r$, but also of other coordinates of the internal space.
We will now give here a simple way to construct non trivial harmonic
functions over the Swann hyperkahler metrics that we have presented.

   First let us notice that all the 4-dimensional quaternion Kahler orbifolds
we have considered possess two commuting Killing vectors, namely
$\partial_{\widetilde{\phi}}$ and $\partial_{\widetilde{t}}$, which
also preserve the one forms $\omega_-^i$ also. Thus these vectors
also preserve the Kahler triplet
$d\overline{J}=d\omega_-+\omega_-\wedge \omega_-$. Consequently they
preserve the hyperkahler triplet (\ref{quato}) for the corresponding
Swann fibration. Such vectors are therefore Killing and
tri-holomorphic (thus tri-hamiltonian) vectors of $g_8$. For any
eight dimensional hyperkahler metric with two commuting
tri-holomorphic Killing vectors there exist a coordinate system in
which it takes the form \cite{Poon} \be\lb{gengibbhawk}
g_8=U_{ij}dx^i\cdot dx^j+ U^{ij}(dt_i+A_i)(dt_j+A_j), \ee being
$(U_{ij}, A_i)$ solutions of the generalized monopole equation
$$
F_{x_{\mu}^i
x_{\nu}^j}=\epsilon_{\mu\nu\lambda}\nabla_{x_{\lambda}^i}U_j,
$$
\be\lb{genmonop}
\nabla_{x_{\lambda}^i}U_j=\nabla_{x_{\lambda}^j}U_i, \ee
$$
U_i=(U_{i1}, U_{i2}),
$$
The hyperkahler form corresponding to (\ref{gengibbhawk}) is
\cite{Poon}
$$
\overline{J}_k=(dt_i+A_i)dx^i_k-U_{ij}(dx^i \wedge dx^j)_k,
$$
From the last expression it follows that the coordinates $dx^j_k$
are defined by means of the relation \be\lb{mch}
dx^{1}_k=i_{\partial_1}\overline{J}_k,\;\;\;\;dx^{2}_k=i_{\partial_2}\overline{J}_k.
\ee The coordinates  $(x^1_i, x^2_i)$ defined by (\ref{mch}) are
known as the momentum maps of the isometries $\partial_1$ and
$\partial_2$. Thus, the generic Gibbons-Hawking form is obtained by
going to the momentum map system.

   In the momentum map system the
11-dimensional supergravity solution reads \be\lb{11sugra2}
g_{11}=H^{-2/3}g_{2,1}+H^{1/3}[U_{ij}dx^i\cdot dx^j+
U^{ij}(dt_i+A_i)(dt_j+A_j)], \ee \be\lb{11sugraF}
F=\pm\omega(E^{2,1})\wedge dH^{-1}, \ee and the harmonic condition
on $H$ is expressed as
$$
U^{ij}\partial_i \cdot \partial_j H=0.
$$
Let us recall that, as a consequence of (\ref{genmonop}), we have
that $\partial_i\cdot\partial_j U_{ij}=0$, which implies that
$U^{ij}\partial_i\cdot\partial_j U_{ij}=0$. This means that any
entry $U_{ij}$ is an harmonic function over the hyperkahler cone.
The matrix $U^{ij}$ is determined by the relation
$U^{ij}=g_8(\partial_i,
\partial_j)$, and the inverse matrix $U_{ij}$ will give us three
independent non trivial harmonic functions for the internal space in
consideration. This give a way to find harmonic functions which are
not only radial, but depends on other coordinates of the cone.

   As an example we can consider the cone $g_8=du^2+u^2g_7$ being
$g_7$ the tri-Sasaki metric corresponding to the AdS-Kerr-Newmann
solution (\ref{azere}). For this cone we have that
$$
U^{\widetilde{t}\widetilde{t}}=u^2(\widetilde{r}-N)^2+  u^2
V(\widetilde{r}), \qquad U^{\widetilde{\phi}\widetilde{\phi}}=4
N^2u^2 V(\widetilde{r})\cos^2\widetilde{\theta}
+(\widetilde{r}^2-N^2)u^2\sin^2\widetilde{\theta}
$$
\be\lb{noche} U^{\widetilde{t}\widetilde{\phi}}=- 2Nu^2
V(\widetilde{r})\cos\widetilde{\theta}
+u^2(\widetilde{r}-N)g(\widetilde{r})\cos\widetilde{\theta}. \ee By
defining $\Delta=U^{\widetilde{t}\widetilde{t}}
U^{\widetilde{\phi}\widetilde{\phi}}-(U^{\widetilde{t}\widetilde{\phi}})^2$
we obtain the following harmonic functions \be\lb{dia}
U_{\widetilde{t}\widetilde{t}}=\frac{U^{\widetilde{\phi}\widetilde{\phi}}}{\Delta},
\qquad U_{\widetilde{\phi}\widetilde{\phi}}=
-\frac{U_{\widetilde{t}\widetilde{t}}}{\Delta}, \qquad
U_{\widetilde{t}\widetilde{\phi}}=\frac{U^{\widetilde{t}\widetilde{\phi}}}{\Delta}.
\ee For the AdS-Kerr-Newman case (\ref{azere2}) we have
$$
U^{\widetilde{t}\widetilde{t}}=u^2\frac{
c(\widetilde{r})d(\widetilde{r})}{\widetilde{r}^2-a^2\cos\widetilde{\theta}^2}
+a^2u^2\frac{f(\widetilde{\theta}) \sin^2\widetilde{\theta}
}{\widetilde{r}^2-a^2\cos^2\widetilde{\theta}} +u^2
e^2(\widetilde{r}, \widetilde{\theta})
$$
$$
U^{\widetilde{\phi}\widetilde{\phi}}=u^2\frac{
c(\widetilde{r})d(\widetilde{r})}{\widetilde{r}^2-a^2\cos\widetilde{\theta}^2}\frac{a^2\sin^4\widetilde{\theta}}{\Xi^2}
+u^2\frac{f(\widetilde{\theta}) \sin^2\widetilde{\theta}
}{\widetilde{r}^2-a^2\cos^2\widetilde{\theta}}\frac{c^2(\widetilde{r})}{\Xi^2}
+u^2\frac{W^2(\widetilde{r}, \widetilde{\theta})}{\Xi^2
e^2(\widetilde{r}, \widetilde{\theta})}
+u^2\frac{f(\widetilde{\theta})
c(\widetilde{r})d(\widetilde{r})}{e^2(\widetilde{r},
\widetilde{\theta})} \frac{\sin^2\widetilde{\theta}}{\Xi^2}
$$
$$
U^{\widetilde{t}\widetilde{\phi}}=u^2 \frac{
c(\widetilde{r})d(\widetilde{r})}{\widetilde{r}^2-a^2\cos\widetilde{\theta}^2}\frac{a\sin^2\widetilde{\theta}}{\Xi}
+au^2\frac{f(\widetilde{\theta}) \sin^2\widetilde{\theta}
}{\widetilde{r}^2-a^2\cos^2\widetilde{\theta}}
\frac{c(\widetilde{r})}{\Xi} +u^2\frac{W(\widetilde{r},
\widetilde{\theta})}{\Xi}
$$
and again, the three functions $U^{ij}/\Delta$ are harmonic
functions over the internal hyperkahler space. Notice that
$\Delta\sim r^4$ and therefore all these harmonic functions depends
on $r$ as $1/u^2$. In the $S^4$ manifold limit we obtain
$$
U_{\widetilde{\phi}\widetilde{\phi}}=-\frac{1}{u^2}\bigg(\frac{1}{\sin^2\widetilde{\theta}
+\sin^2\widetilde{\rho}\cos^2\widetilde{\theta}
-\sin^4\widetilde{\rho}\cos^2\widetilde{\theta}}\bigg),
$$
\be\lb{arm2} U_{\widetilde{\phi}\widetilde{t}}=\frac{1}{u^2}\bigg(\frac{\sin^2\widetilde{\theta}
+\sin^2\widetilde{\rho}\cos^2\widetilde{\theta}}{\sin^2\widetilde{\theta}
+\sin^2\widetilde{\rho}\cos^2\widetilde{\theta}
-\sin^4\widetilde{\rho}\cos^2\widetilde{\theta}}\bigg) \ee
$$
U_{\widetilde{t}\widetilde{t}}=\frac{1}{u^2}\bigg(\frac{\sin^2\widetilde{\rho}\cos\widetilde{\theta}}{\sin^2\widetilde{\theta}
+\sin^2\widetilde{\rho}\cos^2\widetilde{\theta}
-\sin^4\widetilde{\rho}\cos^2\widetilde{\theta}}\bigg)
$$
These harmonic functions provide non $AdS_4$ horizon limits.

  We have constructed the harmonic functions
we were looking for, but it will be instructive to cast our Swann
fibrations into the Gibbons-Hawking form (\ref{gengibbhawk}).
We will do this for the cone $g_8=du^2+u^2g_7$, being $g_7$ the
tri-Sasaki fibration corresponding to the AdS-Taub-Nut
space, which was given in (\ref{azere}). Let us recall that once
the matrix $U^{ij}$ (and its inverse $U_{ij}$) together with
the one forms $A_i$ and the momentum map system $x^i$ are found, the procedure
is finished. The matrix $U^{ij}$ and its inverse were found
already in (\ref{noche}) and (\ref{dia}). Let us turn to the task to
find the momentum map system. From the expression of the Kahler form
(\ref{quato}) we see that the hyperkahler triplet is given in
quaternionic form by
$$
\widetilde{\overline{J}}=u\overline{J}\overline{u}+(du + u
\omega_{-})\wedge \overline{(du + u \omega_{-})},
$$
being the components of the pure quaternionic one form $\omega_-$
given by (\ref{rn}) and the components of the hyperkahler triplet
$\overline{J}$ given by
$$
\overline{J}_i=-e^1\wedge e^2+e^3\wedge e^4,
$$
\be\lb{troplot} \overline{J}_i=e^1\wedge e^3-e^4\wedge e^2, \ee
$$
\overline{J}_i=e^1\wedge e^4-e^2\wedge e^3,
$$
where the soldering forms $e^i$ for AdS-Taub-Nut metric are
$$
e^1=\frac{d\widetilde{r}}{\sqrt{V(\widetilde{r})}}, \qquad
e^2=-\sqrt{\widetilde{r}^2-N^2}d\widetilde{\theta}
$$
$$
e^3=\sqrt{V(\widetilde{r})}(d\widetilde{t}-2N \cos\widetilde{\theta}
d\widetilde{\phi}), \qquad
e^4=-\sqrt{\widetilde{r}^2-N^2}\sin\widetilde{\theta}
d\widetilde{\phi}.
$$
We need to find the solution of the system (\ref{mch}) being
$t_1=\widetilde{t}$ and $t_2=\widetilde{\phi}$. For an arbitrary
vector field $X$ we have that \be\lb{contro}
i_X\widetilde{\overline{J}}=u(\overline{J}(X)+\omega_{-}(X)
\omega_{-}-\omega_{-}\omega_{-}(X))\overline{u}+ u
\omega_{-}(X)d\overline{u}+du \omega_{-}(X)\overline{u}. \ee By
applying the contraction formula (\ref{contro}) with
$X=\partial_{\widetilde{t}}$ or $X=\partial_{\widetilde{\phi}}$ and
taking into account the expressions (\ref{rn}) and (\ref{troplot})
we will obtain a complicated expression for $dx^{\widetilde{t}}$ and
$dx^{\widetilde{\phi}}$. After a lengthy calculation it is found a
compact expression for the momentum maps, which in quaternion
form reads \be\lb{momo}
x^1=u\;[\;(\widetilde{r}-N)\;J\;]\;\overline{u}, \qquad
x^2=u\;[\;g(\widetilde{r})\cos\widetilde{\theta}+\sqrt{(\widetilde{r}+N)V(\widetilde{r})}\sin
\widetilde{\theta}\;K\;]\;J\;\overline{u}. \ee We can express
(\ref{momo}) in pure angular fashion by parameterizing $u$ as in
(\ref{sos}).

  The next step is to find the one forms $A_i$, which are determined by \be\lb{conditions}
A_{i}=U_{ij}\;g_8(\frac{\partial}{\partial t^j},\cdot). \ee The only
quantity to calculate in (\ref{conditions}) is
$g_8(\frac{\partial}{\partial t^j},\cdot)$, the matrix $U_{ij}$ is
defined by (\ref{dia}) and (\ref{noche}). The result is finally
$$
g_8(\frac{\partial}{\partial \widetilde{t}},\cdot)=-4\;N\;u^2
\;V(\widetilde{r})\;\cos\widetilde{\theta}
d\widetilde{\phi}+2\;u^2\;(\widetilde{r}-N)\bigg(\;g(\widetilde{r})\cos\widetilde{\theta}
d\widetilde{\phi}-\sigma_2\;\bigg)
$$
$$
g_8(\frac{\partial}{\partial \widetilde{\phi}},\cdot)=-4\;N\;u^2\;V(\widetilde{r})\;\cos\widetilde{\theta}
\;d\widetilde{\phi}+2\;u^2\;\sqrt{(\widetilde{r}+N)V(\widetilde{r})}\;\sin
\widetilde{\theta}\;\sigma_1
$$
$$
+
2\;u^2\;g(\widetilde{r})\;\cos\widetilde{\theta}\;\bigg(\;(\widetilde{r}-N)\;d\widetilde{t}-\sigma_2\bigg).
$$
 We have calculated all the relevant quantities and the procedure is
finished.

\section{Discussion}

  We have found a family of tri-Sasaki metrics fibered over the
an euclidean quaternion Kahler version of the Plebanski-Demianski
metrics. There exist in the literature examples of 5-dimensional
Einstein-Sasaki spaces related to the Plebanski-Demianski metrics
\cite{Mateos4}-\cite{Mateos9}, but the metrics that we are
presenting are different, they are indeed 7-dimensional. We have
also presented an squashed version with weak $G_2$ holonomy. We have
considered different limits of the quaternion kahler basis, which in
the general case are orbifolds, but in certain limiting cases reduce
to CP(2) or $S^4$, the two compact quaternion Kahler spaces which
are manifolds. We have also extended our solutions to certain
11-dimensional supergravity backgrounds, some with $AdS_4$ near
horizon limit and others that not.

   We would like to remark that the examples that we have presented
here are compact with $T^3$ isometry and therefore are suitable for
applications related to marginal deformations of field theories
\cite{Lunin}. This will be part of a forthcoming paper.

\end{document}